\documentstyle{mn}

\include {epsf}

\begin{document}

\def\lesssim{\mathrel{\hbox{\rlap{\hbox{\lower4pt\hbox{$\sim$}}}\hbox{$<$}}}}

\def\gtrsim{\mathrel{\hbox{\rlap{\hbox{\lower4pt\hbox{$\sim$}}}\hbox{$>$}}}}

\def\msun{$M_{\odot}$~}

\def\teff{$T_{\rm eff}$~}

\def\ll_lsun{$\log{L/L_{\odot}}$~}

\def\masa_msun{$M/M_{\odot}$~}

\def\m_mstar{$M/M_{*}$~}

\def\mean#1{{\langle}#1{\rangle}}

\title{Grids of  white  dwarf evolutionary models  with masses
from M= 0.1 \msun to M= 1.2 \msun}

\author[O. G. Benvenuto \& L. G. Althaus]
{O. G. Benvenuto\thanks{Member  of  the  Carrera  del Investigador
Cient\'{\i}fico, Comisi\'on  de Investigaciones  Cient\'{\i}ficas
de   la   Provincia   de   Buenos   Aires,   Argentina.    Email:
obenvenuto@fcaglp.fcaglp.unlp.edu.ar} and
L. G. Althaus\thanks{Fellow of the Consejo Nacional de
Investigaciones Cient\'{\i}ficas y T\'ecnicas
Argentina.  Email:  althaus@fcaglp.fcaglp.unlp.edu.ar}\\
Facultad de Ciencias Astron\'omicas y
Geof\'{\i}sicas, Paseo del Bosque S/N,
(1900) La Plata, Argentina}

\maketitle

\begin{abstract}  We  present  detailed evolutionary calculations
for carbon - oxygen - and  helium - core white dwarf models  with
masses  ranging  from  $M=  0.1$  to  $M=  1.2$  \msun  and   for
metallicities $Z=$ 0.001 and $Z=$  0. The sequences cover a  wide
range  of  hydrogen  envelopes  as  well.  We have taken finite -
temperature effects  fully into  account by  means of  a detailed
white  dwarf  evolutionary  code,  in  which  updated   radiative
opacities and equations of state for hydrogen and helium  plasmas
are considered.  The energy  transport by  convection is  treated
within the formalism of the full - spectrum turbulence theory, as
given  by  the  self  -  consistent  model of Canuto, Goldman and
Mazzitelli. Convective mixing, crystallization, hydrogen  burning
and neutrino energy losses are taken into account as well.

The set of models presented  here is very detailed and  should be
valuable,   particularly   for   the   interpretation   of    the
observational data on low - mass white dwarfs recently discovered
in  numerous  binary  configurations  and  also  for  the general
problem of  determining the  theoretical luminosity  function for
white dwarfs. In this  context, we compare our  cooling sequences
with  the  observed  white  dwarf  luminosity  function  recently
improved by Leggett,  Ruiz \& Bergeron  and we obtain  an age for
the Galactic disc  of $\approx 8\  Gyr$. Finally, we  applied the
results of this paper to derive stellar masses of a sample of low
- mass white dwarfs.

\end{abstract}


\section{introduction}

Numerous  observations  carried  out  over recent  years have
presented strong  evidence that  low -  mass, helium  white dwarf
stars are the  product of the  evolution of certain  close binary
systems. Indeed, low  - mass white  dwarfs have been  detected in
binary  systems  containing,  for  instance,  another white dwarf
(Marsh 1995;  Marsh, Dhillon  \& Duck  1995; Marsh  \& Duck 1996;
Moran, Marsh \& Bragaglia 1997),  millisecond pulsar (Lundgren
et al. 1996; see also Backer 1998) or a yellow giant (Landsman et
al. 1997). In  particular, Moran et  al. (1997) found  the binary
system WD 0957 - 666 (consisting of two low - mass white  dwarfs)
to have an orbital period of  only 1.46 h, which is short  enough
for the binary to merge within  only $2.0 \times 10 ^8$ yr.  Very
recently, Edmonds et al. (1998)  have reported the presence of  a
candidate helium white dwarf in the globular cluster NGC 6397. On
theoretical grounds, recent  population models of  close binaries
(Iben, Tutukov \& Yungelson  1997) suggest a high  probability of
discovering helium white dwarfs in close binaries.

Detailed  evolutionary  models  of  low  -  mass white dwarfs may
provide valuable information not  only on the white  dwarf itself
but also on the companion  object and even on the  past evolution
of the system (see, for instance, Burderi, King \& Wynn 1996  and
Hansen \& Phinney  1998a). In this  regard, the analysis  carried
out, notably  by van  Kerkwijk, Bergeron  \& Kulkarni  (1996), is
worth mentioning. Indeed,  from spectroscopic data  inferred from
its low - mass white dwarf companion combined with a  theoretical
mass - radius relation for  the white dwarf, these authors  found
the mass of the pulsar PSR J1012 + 5307 to be between 1.5 and 3.2
\msun. Needless to say,  detailed models of helium  white dwarfs,
together  with  further  observations,  are  needed  in  order to
achieve a more precise determination of the pulsar mass and hence
to  constrain  the  equation  of  state  at  the  high  densities
appropriate  for  neutron  stars.  In  addition,  an  independent
determination  of  the  age  of  many  millisecond pulsars can be
inferred from  the study  of the  cooling of  their helium  white
dwarf companions, which is valuable for understanding the  nature
and  origen  of  such  systems.  Another  strong  motivation  for
constructing improved white  dwarf evolutionary sequences  is the
fact that,  thanks to  the {\it  Hubble Space  Telescope}, it has
been possible to  detect the low  - luminosity tail  of the white
dwarf population in  globular clusters. Accordingly,  white dwarf
evolutionary tracks would provide an independent way of measuring
the age and distance of  such clusters (see, e.g., Richer  et al.
1995; Von Hippel, Gilmore \& Jones 1995 and Renzini et al. 1996).

In view  of these  considerations, we  present in  this paper new
grids of white dwarf  evolutionary models for different  hydrogen
envelopes and stellar  masses. The emphasis  is placed mainly  on
low - mass, helium white dwarfs, the detailed study of which  has
recently began to be undertaken. As a matter of fact, Althaus  \&
Benvenuto (1997a) and Benvenuto \& Althaus (1998) carried out  an
analysis  of  the  structure  and  evolution  of low - mass white
dwarfs  based  on  a  updated  physical  description, such as new
opacities and  equations of  state, and  the employment  of a new
convection model  more physically  sound than  the mixing  length
theory. In a still more  recent study, Hansen \& Phinney  (1998a)
presented evolutionary  calculations for  these objects  as well.
However, the evolutionary sequences for their more massive models
do not converge to the Hamada \& Salpeter (1961) predictions  for
zero - temperature, pure - helium configurations, thus  resulting
in models with underestimated surface gravities. This can be seen
from  figure  16  of  Hansen  \&  Phinney  (1998a). Note that the
surface gravity  for their  more massive  models with  a hydrogen
envelope of $M_{\rm  H}/$\msun= $10^{-6}$ is  substantially lower
than the Hamada - Salpeter  values. Such a discrepancy cannot  be
attributed to the hydrogen layer, since a thin  hydrogen envelope
introduces a very small correction to the stellar radius of  zero
- temperature,  pure helium models  (see Benvenuto  \& Althaus
1998).

By contrast,  the study  of evolution  of carbon  - oxygen  white
dwarfs have captured the interest of numerous investigators  such
as Lamb \&  Van Horn (1975),  Iben \& Tutukov  (1984), Koester \&
Sch\"onberner  (1986),  D'Antona  \&  Mazzitelli (1989), Tassoul,
Fontaine  \&  Winget  (1990),  Wood  (1992), Benvenuto \& Althaus
(1997) and Althaus  \& Benvenuto (1998).  In particular, Iben  \&
Tutukov (1984) were the first in computing evolutionary models of
white dwarfs with hydrogen burning, showing that hydrogen burning
in cooling white dwarfs could be an important energy source.

With  the  calculations that we  present  here  we amply extend
those
presented  in  Althaus  \&  Benvenuto  (1997a)  and  Benvenuto \&
Althaus  (1998),  in  which  the  effects of convection, neutrino
losses  and  different  hydrogen  envelopes  on the structure and
evolution of helium white dwarfs were carefully analysed (we
should
mention that  the results shown  in Benvenuto \&  Althaus 1998
correspond to  a metallicity  of $Z  \approx$ 0  and not  to $Z=$
0.001,  as  stated  in  that  work).  Furthermore,  we extend our
calculations to the  case of more  massive carbon -  oxygen white
dwarfs. Our  grid for  carbon -  oxygen models  is likewise  very
detailed,  which  may  be  of  relevance  in  the  study  of, for
instance,  the  general  problem  of  determining the theoretical
white dwarf luminosity function and the assessment of the age  of
the Galactic disc.  This subject has  been recently addressed  by
Leggett, Ruiz \& Bergeron  (1998), who have greatly  improved the
determination of the observed luminosity function for cool  white
dwarfs. In  this regard,  we shall  derive theoretical luminosity
functions from our cooling sequences in order to compare with the
Leggett et al. observational data.

The  results  presented  here  constitute  a  very  detailed and
updated  set  which  will  be  suitable,  for  instance,  for the
interpretation of recent and forthcoming observational data about
low -  mass white  dwarfs in  close binary  systems. Finally,  we
applied  our  evolutionary  models  with  helium  cores to derive
stellar masses of a sample of low - mass white dwarfs.

\section{Computational Details} \label{sec_physics}

The  evolutionary  sequences  have  been  obtained  with the same
evolutionary code and input  physics we employed in  our previous
works  on  white  dwarf  evolution,  and  we  refer the reader to
Althaus  \&  Benvenuto  (1997a,  1998)  and  Benvenuto \& Althaus
(1998) as well as to the references cited therein for details. In
what follows we restrict ourselves to a few brief comments.

The  code  has  been  written  following  the method presented by
Kippenhahn, Weigert \& Hofmeister (1967) for calculating  stellar
evolution.  In  particular,  to  specify  the  surface   boundary
conditions we  perform three  envelope integrations  (at constant
luminosity) from photospheric starting values inward to a fitting
mass   fraction   $M_{1}/M   \approx   10^{-16}$,  where  $M_{1}$
corresponds to the first Henyey  mass shell and $M$ is  the total
mass of the white dwarf model.  In our code the value of  $M_{1}$
is automatically  changed over  the evolution  so as  to keep the
thickness of the envelope as small as possible. This provides  an
accurate  description  of  the  outer  layers  of our white dwarf
models.  The  interior  integration  is  treated according to the
standard  Henyey  technique  as  described  by  Kippenhahn et al.
(1967).
\begin{table*}
 \centering
\begin{minipage}{140mm}
\caption{Available white dwarf evolutionary sequences with
hydrogen envelopes}
\begin{tabular}{@{}cccccccc@{}}
Sequence & $T^i_{\rm eff}$ &
Sequence & $T^i_{\rm eff}$ &
Sequence & $T^i_{\rm eff}$ &
Sequence & $T^i_{\rm eff}$ \\
 He151e3.z3  &  8.12 & He151e4.z3 &  8.97 &   He151e6.z3  &  9.53 & He151e8.z3 &  9.97   \\
 He152e3.z3  &  7.80 & He152e4.z3 &  8.80 &   He154e3.z3  &  7.61 & He154e4.z3 &  8.62   \\
 He201e3.z3  & 11.87 & He201e4.z3 & 13.96 &   He201e6.z3  & 14.75 & He201e8.z3 & 14.97   \\
 He202e3.z3  & 11.93 & He202e4.z3 & 12.71 &   He204e3.z3  & 11.38 & He204e4.z3 & 12.71   \\
 He251e3.z3  & 17.28 & He251e4.z3 & 19.04 &   He251e6.z3  & 20.24 & He251e8.z3 & 21.35   \\
 He252e3.z3  & 16.41 & He252e4.z3 & 18.77 &   He254e4.z3  & 18.32 & He301e3.z3 & 22.14   \\
 He301e4.z3  & 23.92 & He301e6.z3 & 25.76 &   He301e8.z3  & 26.15 & He302e3.z3 & 21.73   \\
 He302e4.z3  & 23.43 & He304e4.z3 & 22.64 &   He351e3.z3  & 23.90 & He351e4.z3 & 24.65   \\
 He351e6.z3  & 25.51 & He351e8.z3 & 25.82 &   He352e4.z3  & 24.00 & He354e4.z3 & 24.02   \\
 He361e3.z3  & 26.14 & He371e3.z3 & 26.57 &   He381e3.z3  & 27.66 & He391e3.z3 & 28.78   \\
 He401e3.z3  & 22.50 & He401e4.z3 & 27.31 &   He401e6.z3  & 28.72 & He401e8.z3 & 30.55   \\
 He402e4.z3  & 26.61 & He404e4.z3 & 25.47 &   He411e3.z3  & 30.52 & He428e4.z3 & 31.10   \\
 He429e4.z3  & 31.28 & He439e4.z3 & 32.00 &   He448e4.z3  & 32.84 & He451e4.z3 & 31.11   \\
 He451e6.z3  & 29.30 & He451e8.z3 & 30.67 &   He452e4.z3  & 29.77 & He454e4.z3 & 27.96   \\
 He458e4.z3  & 34.20 & He468e4.z3 & 35.60 &   He501e4.z3  & 36.62 & He501e6.z3 & 39.18   \\
 He501e8.z3  & 38.22 & He502e4.z3 & 37.07 &   He504e4.z3  & 35.85 &             &        \\
OC501e4.z3 &  57.91 & OC501e6.z3 &  60.78 & OC501e8.z3 &  61.90 & OC50e10.z3 &  61.31 \\
OC50e12.z3 &  61.42 & OC601e4.z3 &  61.07 & OC601e6.z3 &  70.69 & OC601e8.z3 &  72.04 \\
OC60e10.z3 &  74.69 & OC60e12.z3 &  74.81 & OC701e4.z3 &  73.47 & OC701e6.z3 &  78.56 \\
OC701e8.z3 &  87.23 & OC70e10.z3 &  87.94 & OC70e12.z3 &  88.07 & OC801e4.z3 &  91.60 \\
OC801e6.z3 &  99.48 & OC801e8.z3 & 101.14 & OC80e10.z3 & 108.68 & OC80e12.z3 & 108.83 \\
OC901e6.z3 & 121.03 & OC901e8.z3 & 122.79 & OC90e10.z3 & 120.61 & OC90e12.z3 & 122.81 \\
OC101e6.z3 & 134.59 & OC101e8.z3 & 136.49 & OC10e10.z3 & 136.19 & OC10e12.z3 & 136.37 \\
OC111e6.z3 & 198.47 & OC111e8.z3 & 196.97 & OC11e10.z3 & 195.85 & OC11e12.z3 & 194.00 \\
He154e3.z0 &  7.70  & He201e4.z0 & 14.05  & He204e3.z0 & 11.37  & He251e3.z0 & 17.01  \\
He251e4.z0 & 18.93  & He252e3.z0 & 16.38  & He301e3.z0 & 20.55  & He301e4.z0 & 23.63  \\
He302e3.z0 & 21.70  & He351e3.z0 & 23.50  & He351e4.z0 & 24.83  & He401e3.z0 & 22.69  \\
He401e4.z0 & 27.14  & He454e4.z0 & 29.49  \\
OC451e4.z0 & 43.28 & OC471e4.z0 & 45.29 & OC501e4.z0 & 58.49 & OC521e4.z0 & 60.08 \\
OC541e4.z0 & 53.71 & OC561e4.z0 & 49.94 & OC581e4.z0 & 49.65 & OC601e4.z0 & 58.86 \\
OC621e4.z0 & 57.49 & OC641e4.z0 & 61.79 & OC661e4.z0 & 62.33 & OC681e4.z0 & 67.93 \\
OC701e4.z0 & 77.50 & OC721e4.z0 & 77.90 & OC741e4.z0 & 79.33 & OC761e4.z0 & 83.65 \\
OC781e4.z0 & 85.30 & OC801e4.z0 & 81.06 & OC821e4.z0 & 81.79 & OC841e4.z0 & 89.26 \\
OC901e4.z0 & 106.71 & OC101e4.z0 & 106.95 & OC111e4.z0 & 95.00 & OC121e6.z0 & 96.65 \\
OC121e8.z0 & 97.27
\end{tabular}
\medskip

This table shows available  evolutionary  sequences  for  white
dwarf models with
a hydrogen envelope.  We use  an abbreviated  notation to
indicate
the core composition,  the stellar mass  in tenths of  solar mass
units,  the  mass  fraction  of  the  hydrogen  envelope  and the
envelope metallicity.  For instance,  He252e3 .z0  stands for  an
evolutionary sequence of 0.25$M_{\sun}$ models with a helium core
composition,  a  hydrogen  envelope  of  $M_{\rm  H}/M=  2 \times
10^{-3}$ and an envelope metallicity of $Z=0$. OC means a  oxygen
- carbon core composition (Note that OC sequences for model  more
massive than 1$M_{\sun}$ are  indicated with the same  notation).
We also  provide a  column ($T^i_{\rm  eff}$) with  the effective
temperature  (in  thousand  K  degrees)  at  which  each sequence
starts.
\end{minipage}
\end{table*}

The constitutive physics of our  code is as detailed and  updated
as possible. Briefly, for the  low - density regime, we  consider
the equation of state of Saumon, Chabrier \& Van Horn (1995)  for
hydrogen  and  helium  plasmas.  The  treatment  for  the  high -
density,  completely  ionized  regime  appropriate  for the white
dwarf  interior  is  based  on  our  own  equation of state. This
includes ionic  and photon  contributions, coulomb  interactions,
partially degenerate electrons, quantum corrections for the  ions
and electron exchange and Thomas - Fermi contributions at  finite
temperature  (see  Althaus  \&  Benvenuto  1997a  for   details).
Radiative opacitites for the  high - temperature regime  ($T \geq
6000$ K) are those of OPAL (Iglesias \& Rogers 1993), whilst  for
lower  temperatures  we  use  the  Alexander  \&  Ferguson (1994)
molecular opacities (or  the Cox \&  Stewart 1970 tabulation  for
pure helium composition). Two extreme values for metallicity have
been considered in the envelope: $Z$= 0 and $Z$= 0.001. We should
mention that, owing to the lack of reliable low - temperature ($T
< 6000 K$) opacities for helium composition, our low - luminosity
models with helium atmospheres  should be regarded with  caution,
particularly their ages. Conductive opacities for the liquid  and
crystalline  phases  and  the  various  mechanisms  of   neutrino
emission relevant  to white  dwarf interiors  are taken  from the
works of Itoh and  collaborators (see Althaus \&  Benvenuto 1997a
for details). We also include in our code the complete network of
thermonuclear reaction rates  for hydrogen burning  corresponding
to the  proton -  proton chain  and the  CNO bi  - cycle. Nuclear
reaction  rates  are  taken  from  Caughlan  \& Fowler (1988) and
$\beta$ - decay rates from Wagoner (1969). Electron screening  is
from Wallace, Woosley \& Weaver (1982). We use an implicit method
of integration to  compute the change  of the following  chemical
species:   $^{1}$H,   $^{2}$H,   $^{3}$He,   $^{4}$He,  $^{7}$Li,
$^{7}$Be,  $^{8}$B,   $^{12}$C,  $^{13}$C,   $^{13}$N,  $^{14}$N,
$^{15}$N, $^{15}$O, $^{16}$O, $^{17}$O and $^{17}$F.

\begin{figure}
\epsfxsize=240pt
\begin{displaymath}
\epsfbox{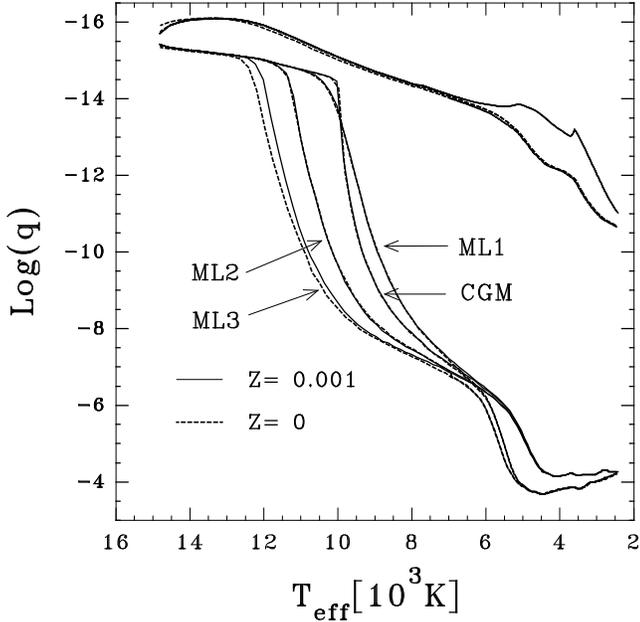}
\end{displaymath}
\caption{The location of  the top and the  base of
the convection zone in terms of the outer mass fraction
$q$  ($q=  1-M_r/M$)  versus $T_{eff}$  for a 0.3 $M_{\odot}$
white  dwarf   model  with   a  hydrogen   envelope  of   $M_{\rm
H}/M=10^{-3}$ according to  different theories of  convection and
metallicities. It is clear that the mixing temperature for models
with thin hydrogen envelopes  will be dependent upon  the assumed
theory  of  convection.  At  low $T_{eff}$ values, the depth
reached by the base of the convection zone is independent of  the
treatment  of  convection.  Note  also  the  deeper  final extent
reached  by  convection  in  the  case  of  metallicity   $Z=0$.}
\label{fig_1}
\end{figure}

Another important feature of  our evolutionary sequences is  that
the energy  transport by  convection is  described by  the full -
spectrum turbulence theory (see  Canuto \& Mazzitelli 1991,  1992
and references  cited therein  for details),  which represents  a
great  improvement  compared  with  the  mixing  length theory of
convection used  thus far  in most  of white  dwarf studies. As a
matter  of  fact,  the  Canuto  \&  Mazzitelli  theory, which has
successfully   passed   a   wide   variety   of   laboratory  and
astrophysical tests (Canuto 1996),  takes into account the  whole
spectrum of turbulent eddies necessary to compute the  convective
flux accurately in the almost inviscid stellar interiors. For the
set of sequences presented in  this paper we have considered  the
recent  improvement  to  this  convection  theory  introduced  by
Canuto, Goldman  \& Mazzitelli  (1996), which  has been  shown to
provide  a  good  agreement  with  recent  observational  data on
pulsating white  dwarfs (Althaus  \& Benvenuto  1997b). The model
presented by Canuto  et al. improves  over its predecesor  (where
the rate of input energy is  given by the linear growth rate)  in
the fact that the  growth rate is computed  as a function of  the
turbulence itself, thus ensuring  a self - consistent  treatment.
At  intermediate  and  low  convective  efficiency , this feature
leads to larger convective fluxes as compared with the Canuto  \&
Mazzitelli (1992)  model. It  is worthwhile  to mention  that the
mass - radius relation and ages corresponding to our white  dwarf
models  are  practically  insensitive  to  the  convection theory
employed. In contrast, the size  of the outer convection zone  in
an  intermediate  effective  temperature  (\teff), evolving white
dwarf  is  strongly  dependent  upon  the  assumed  treatment  of
convection. Hence, a trustworthy  model of stellar convection  is
to  be  employed  to  get  reliable  \teff  values  at which thin
hydrogen envelopes mix with the underlying helium (see  Benvenuto
\& Althaus 1998). In this context, our calculations represent  an
improvement over previous white dwarf studies based on the mixing
length theory  of convection.  To clarify  this point  better, we
show in  Figure~\ref{fig_1} the  behaviour of  the evolving outer
convection zone in terms of \teff for our 0.3 \msun model with  a
thick hydrogen envelope. In addition to Canuto et al.'s  results,
we include in  the figure the  predictions given by  the ML1, ML2
and ML3 versions of the mixing length theory amply used in  white
dwarf studies (see, e.g., Tassoul et al. 1990). It is clear  that
the mixing \teff for models with thin hydrogen envelopes  depends
upon the convection theory. Another observation that we can  make
from this figure is that the base of the convection zone (for the
case $Z=0$) ultimately  reaches a final  extent at an  outer mass
fraction  of  $2  \times  10^{-4}$  ($6  \times  10^{-5}$ \msun),
irrespective of the  treatment of convection.  This result is  in
good agreement with the predictions of Hansen \& Phinney  (1998a)
for  the  same  model.  Note  also  the  larger  final  extent of
convection zone  for lower  metallicity, which  as we  shall see,
gives rise to considerable differences in the evolutionary  times
at low luminosities.

\begin{figure}
\epsfxsize=240pt
\begin{displaymath}
\epsfbox{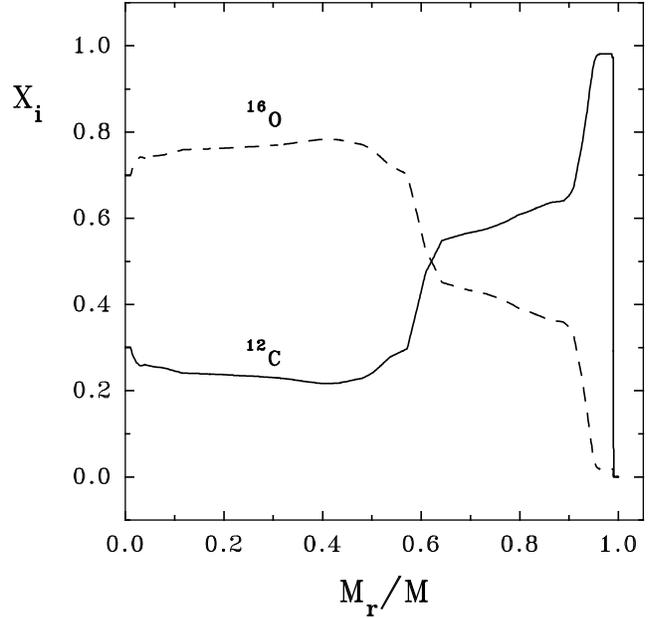}
\end{displaymath}
\caption{  Chemical  profile  for  our carbon -
oxygen  core  models  versus  the  fractional  mass.  Solid lines
correspond  to  carbon  and  medium  dashed  lines  to   oxygen.}
\label{fig_2}
\end{figure}

Our white dwarf initial  models of different masses  and hydrogen
envelopes   have   been   obtained   following   the   artificial
evolutionary procedure described by Benvenuto \& Althaus  (1998).
The carbon - oxygen core  models all have the same  core chemical
composition profile  shown in  Figure~\ref{fig_2}. This  chemical
profile was calculated by  D'Antona \& Mazzitelli (1989)  for the
progenitor evolution of a 0.55  \msun white dwarf. We adopt  this
profile for all of our models,  in spite of the changes that  are
expected  to  occur  for  more  massive  models  as  a  result of
differences  in  the  evolution  in  progenitor objects. We would
need, in order to improve this assumption, detailed  calculations
of the  pre -  white dwarf  evoltution of  these objects.  To our
knowledge, such  calculations are  not available.  Because of the
fact that the  mass of the  hydrogen envelope in  white dwarfs is
poorly  constrained  by  theoretical  calculations  of  the pre -
evolution of these  objects, particularly in  the case of  helium
white dwarfs where the uncertainties regarding the mass  exchange
episodes are  more severe,  we decide  to treat  the mass  of the
hydrogen  envelope  as  essentially  a  free  parameter.  It   is
worthwhile to mention  that our evolving  low - mass  white dwarf
models should be  considered as evolutionary  stages that can  be
asymptotically  reached  by  helium  white  dwarfs resulting from
close binary evolution. In this  study we have not computed  such
binary evolution, and we refer the reader to Iben \& Livio (1993)
for a review. Needless to  say, the starter model choice  affects
the {\it initial}  evolution of all  of our models,  particularly
the age (see Althaus \& Benvenuto 1997a for details).

In closing, we have included  in our calculations the release  of
latent heat during crystallization (see Benvenuto \& Althaus 1997
for details) and convective mixing.

\section{THE GRIDS OF WHITE DWARF MODELS} \label{sec_resu}

In this section we comment on the most important features of  the
grids.  We  computed  evolutionary  sequences with masses ranging
from $M= 0.1$ \msun to $M= 1.2$ \msun and metallicity $Z=$  0.001
and $Z=$  0. For  models with  $M <  0.5$ \msun  we assume a pure
helium core  and for  models with  $M >  0.45$ \msun  we assume a
carbon  -  oxygen  core  with  the  chemical  profile  of Figure~
\ref{fig_2}.  We  also  vary  the  mass  of the hydrogen envelope
$M_{\rm H}$ within  the range $10^{-12}  \leq M_{\rm H}/M  \leq 4
\times 10^{-3}$ and the mass of the helium layer $M_{\rm He}$ (in
the case of carbon - oxygen white dwarfs) is taken to be  $M_{\rm
He}/M  =  10^{-2}$.  In  Tables  1  and  2  we summarize the main
characteristics of all  of our available  evolutionary sequences.
As stated earlier, models  have been calculated in  the framework
of Canuto et al. (1996)  theory of convection. We have  used OPAL
opacity calculations supplemented with the Alexander \&  Ferguson
(1994)  molecular  opacities  (or  with  the  Cox \& Stewart 1970
tabulation for helium composition) for low temperatures. In Table
1 and 2 we also give the \teff values at which each  evolutionary
sequence starts.  In this  regard, we  emphasise once  again that
model ages corresponding to  the {\it first} stages  of evolution
are  meaningless  because  they  are  strongly  affected  by  the
procedure  we  use  to  generate  the initial models. At advanced
ages, however,  this is  no longer  relevant and  age values  are
meaningful. The  sequences have  been evolved  down to  a stellar
luminosity \ll_lsun= -5. \begin{table*}
 \centering
\begin{minipage}{140mm}
\caption{Available white dwarf evolutionary sequences without a
hydrogen envelope}
\begin{tabular}{@{}cccccccc@{}}
Sequence & $T^i_{\rm eff}$ &
Sequence & $T^i_{\rm eff}$ &
Sequence & $T^i_{\rm eff}$ &
Sequence & $T^i_{\rm eff}$ \\
 He0990.z3 &  6.62 &  He1039.z3 &  7.00 &  He1091.z3 &  7.45 &  He1146.z3 &  7.93 \\
 He1203.z3 &  8.45 &  He1263.z3 &  9.02 &  He1326.z3 &  9.62 &  He1393.z3 & 10.26 \\
 He1462.z3 & 10.99 &  He1500.z3 & 11.59 &  He1575.z3 & 12.17 &  He1654.z3 & 12.90 \\
 He1736.z3 & 13.51 &  He1823.z3 & 14.08 &  He1914.z3 & 14.82 &  He2010.z3 & 15.85 \\
 He2110.z3 & 17.26 &  He2216.z3 & 18.27 &  He2327.z3 & 20.01 &  He2443.z3 & 21.42 \\
 He2565.z3 & 22.83 &  He2693.z3 & 24.61 &  He2829.z3 & 25.29 &  He2970.z3 & 26.70 \\
 He3118.z3 & 26.44 &  He3274.z3 & 26.77 &  He3438.z3 & 27.90 &  He3610.z3 & 28.07 \\
 He3790.z3 & 30.09 &  He3979.z3 & 29.63 &  He4179.z3 & 30.60 &  He4388.z3 & 32.49 \\
 He4607.z3 & 34.63 &  He4837.z3 & 37.30 &  He5079.z3 & 37.69 &            &       \\
OC5000.z3 &  57.61 & OC5200.z3 &  47.60 & OC5400.z3 &  51.96 & OC5600.z3 &  54.21 \\
OC5800.z3 &  59.70 & OC6000.z3 &  78.72 & OC7000.z3 &  78.59 & OC8000.z3 &  93.85 \\
OC9000.z3 &  94.82 & OC1000.z3 & 101.57 & OC1100.z3 & 107.97 & OC1200.z3 & 100.01
\end{tabular}
\medskip

This table shows available evolutionary  sequences for  white
dwarf
models  without  a  hydrogen  envelope.  We  use  an  abbreviated
notation to indicate  the core composition,  the stellar mass  in
tenths  of  solar  mass  units  and the envelope metallicity. For
instance,  He3118.z3  stands  for  an  evolutionary  sequence  of
0.3118$M_{\sun}$ models  with a  helium core  composition and  an
envelope metallicity of $Z=0.001$. OC means oxygen - carbon  core
composition. (Note that OC sequences for model more massive  than
1$M_{\sun}$  are  indicated  with  the  same  notation).  We also
provide a column ($T^i_{\rm eff}$) with the effective temperature
(in thousand K degrees) at which each sequence starts.
\end{minipage}
\end{table*}

\begin{figure}
\epsfxsize=240pt
\begin{displaymath}
\epsfbox{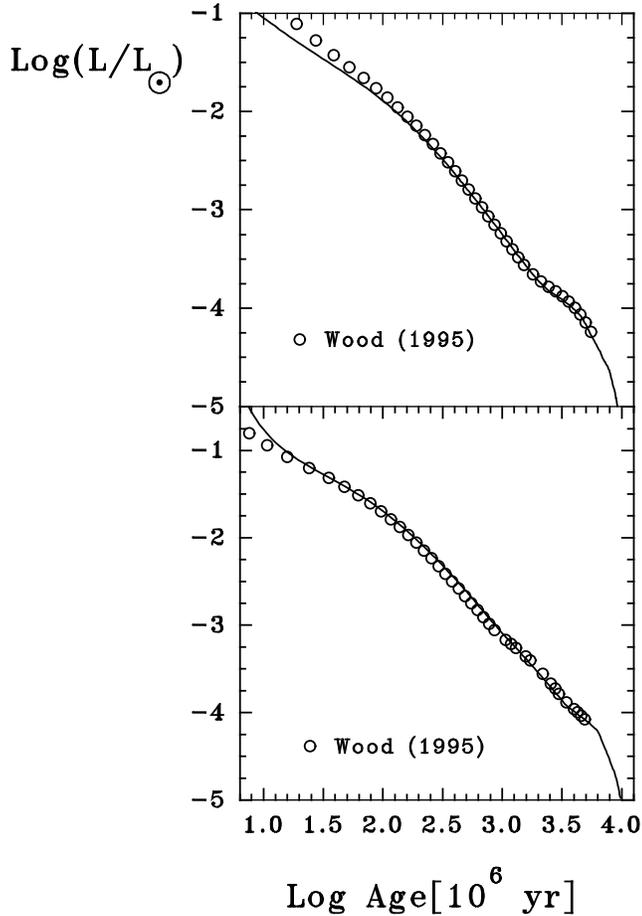}
\end{displaymath}
\caption{  Age -  surface luminosity  relation for
0.5 (upper panel) and 0.7 (lower panel) \msun  white dwarf
models  with  pure  oxygen  cores  and  hydrogen envelope mass of
$M_{\rm H}/M$= $10^{-4}$. We compare our results (solid lines) to
those given by the Wood (1995) models (open circles) having the
same
stellar   mass   and   chemical   stratification   as  ours.  The
calculations   are   for   a   metallicity   of  $Z=0$.  At  high
luminosities, the discrepancy between the two set of calculations
are  due  to  the  different  procedure  employed to generate the
initial models.} \label{fig_3}
\end{figure}

We begin by examining the time evolution of our models. From  the
point of view of an age  determination of the disc of our  Galaxy
from the observed space density of white dwarfs, the evolutionary
times of white dwarfs as  a function of mass represent  obviously
an important issue (see Wood 1992 and references cited  therein).
In this  regard, we  feel it to  be valuable  to compare our
cooling
curves against those published  by other authors. We  elect those
computed by Wood  (1995) for  models with pure  oxygen core on
the basis of OPAL radiative  opacities. To this end, we
have computed
additional sequences for  models with oxygen  cores and with  the
same  outer  layer  chemical  stratification  and  metallicity as
considered  by   Wood.  The   comparison  is   shown  in  Figure~
\ref{fig_3}  for  0.5  and  0.7  \msun  models. Note the good
agreement  between  the  two  set  of  calculations.  At very low
luminosities and  especially for  more massive  models than those
considered  in  Figure~\ref{fig_3},  some  divergency appears as
a
result  in  part  of  the  different  set  of  low  - temperature
radiative opacities employed.

As  we  mentioned,  white  dwarf  evolutionary  times represent a
powerful tool for constraining the age of the disc of our Galaxy.
Indeed, the existence of an abrupt falloff in the observed  white
dwarf luminosity function  (see Liebert, Dahn  \& Monet 1988  and
earlier references cited therein)  has been interpreted in  terms
of a finite age of the disc of the Galaxy (D'Antona \& Mazzitelli
1978). By fitting the  observations with theoretical white  dwarf
luminosity  functions,  this  interpretation  was  quantitatively
explored by numerous investigators such as Winget et al.  (1987),
Iben \& Laughlin (1989) and Wood (1992). Recently, Leggett et al.
(1998)  have  substantially  improved  the  determination  of the
observed luminosity  function for  cool white  dwarfs. To compare
with observations, we constructed integrated luminosity functions
from  our  evolutionary  sequences.  To  this  end, we follow the
treatment presented in Iben \& Laughlin (1989). Specifically, the
space density of  white dwarfs per  unit of $\ell$,  $\ell \equiv
log (L/L_{\sun})$, is calculated from

\begin{equation}
{{dn}\over{d \ell}}= -\psi_o\
\int^{m_s}_{m_i}
{\phi(m)\ ({{{\partial t_{cool}}\over{\partial
\ell}}})_M\ dm}. \label{eq:lf}
\end{equation}

\begin{figure}
\epsfxsize=240pt
\begin{displaymath}
\epsfbox{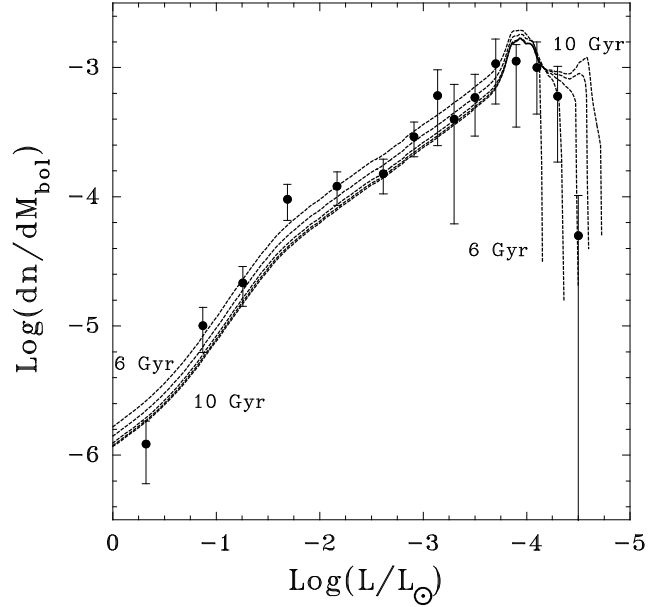}
\end{displaymath}
\caption{  Theoretical  white  dwarf   luminosity
functions (dotted  lines) corresponding  to our  carbon -  oxygen
core,  white dwarf  models with  a hydrogen  envelope mass  of
$M_{\rm H}/M$= $10^{-4}$ and metallicity $Z=0$. The curves, which
correspond to assumed disc ages of 6 - 10 $Gyr$ (at intervals  of
1 $Gyr$), are  compared to the  observational data of  Leggett et
al. (1998) and  they have been  normalized to the  observed white
dwarf space density of 0.00339 stars per cubic parsec. Note  that
the best fit to the dimmest white dwarfs observed correponds to a
disc age of approximately 8 $Gyr$. }
\label{fig_4}
\end{figure}

Here, $\phi(m)$ is  the Salpeter initial  mass function of  white
dwarf progenitors with stellar mass $m$ (which predicts that  the
created star  distribution is  proportional to  $1/m^{2.35}$) and
$t_{cool}$ is  the white  dwarf cooling  time at  a given $\ell$,
which is a function of the white dwarf mass $M$. $m_i$ and  $m_s$
denote respectively the minimum and the maximum mass of the  main
sequence stars which contribute to the white dwarf space  density
at  $\ell$.  We  take  $m_{s}  \approx  8$  \msun (Wood 1992) and
$m_{i}$  is  obtained  by  solving  the  equation  $t_{MS}(m)   +
t_{cool}(\ell,M)= t_{d}$, where $t_{d}$ is the assumed disc  age.
The pre - white dwarf evolutionary times $t_{MS}(m)$ are those of
Iben \& Laughlin (1989). As far as the initial($m$) -  final($M$)
mass  relation  is  concerned,  we  use an exponential model: $M=
0.40\   e^{0.125\   m}$   (Wood   1992).   In  deriving  Equation
(\ref{eq:lf}), the star formation rate $\psi_o$ has been  assumed
to  be  constant.  Finally,  for  each of the selected luminosity
values, we calculate $\partial t_{cool}/\partial \ell$ at a given
$M$  by   using  linear   interpolation  between   the  $\partial
t_{cool}/\partial \ell$ values of the sequences that bracket $M$.
The resulting luminosity functions for  assumed disc ages of 6  -
10 $Gyr$ are shown in Figure~\ref{fig_4} (in Figure~ \ref{fig_4},
the luminosity  functions have  been converted  into intervals of
bolometric magnitude $M_{bol}$). It is worth mentioning that  all
of our theoretical  curves have been  normalized to the  observed
space density of 0.00339  white dwarfs per cubic  parsec (Leggett
et al. 1998). The best  fit to the coolest white  dwarfs observed
is obtained for assumed disc  ages of $\approx 8\ Gyr$,  which is
in agreement with the ages quoted by Leggett et al. on the  basis
of the Wood (1995) cooling sequences.

It is worthy of comment that the age of helium white dwarf models
depends  on  the   mass  of  the   hydrogen  envelope.  This   is
particularly true for models with very thick hydrogen  envelopes,
for which hydrogen burning contributes substantially to the total
luminosity, thus leading to a delay in cooling even down to  very
low  \teff.  Unfortunately,  the  maximum  mass  of  the
hydrogen envelope  resulting from  binary evolution  is still  an
open  question.  Evidence  favouring  ``thin  ''  envelopes  was
presented by Iben \& Tutukov (1986) from self - consistent binary
evolutionary calculations. Indeed,  these authors found  that the
hydrogen envelope remaining  at the top  of their 0.3  \msun
remnant  after  shell  flash  episodes  is  too small ($M_{\rm H}
\approx  1.4  \times  10^{-4}$  \msun)  to sustain any further
nuclear burning. Needless to  say, a larger hydrogen remnant
would
lead  to  longer  evolutionary  times.  In  the  context  of  age
determinations  of  millisecond  pulsars  with helium white dwarf
companions, this fact is a clearly important one to be  taken
into  account.  To  place  this  assertion on a more quantitative
basis, we consider the pulsar PSR J1012+5307. The surface gravity
and \teff for its low - mass helium white dwarf companion
have been  determined to  be log  $g= 6.75  \pm 0.07$ and \teff=
$8550 \pm 25 K$ (van Kerkwijk et al. 1996). At \teff
$\approx 8500 K$,  we find that  our 0.21 \msun  helium white
dwarf model with  $M_{\rm H}/M= 2  \times 10^{-3}$ has  log $g=$
6.82 and  age 0.44  $Gyr$, in  good agreement  with the Hansen \&
Phinney (1998b) predictions.  The same fit  to the \teff
and  gravity  values  would  be  achieved with a 0.213 \msun
model with $M_{\rm  H}/M \approx 6  \times 10^{-3}$. However,  in
this case hydrogen  burning supplies 70  per cent of  the surface
luminosity and the model age becomes as high as 0.9 $Gyr$.
\begin{table*}
 \centering
\begin{minipage}{140mm}
\caption{Masses for low mass white dwarfs}
\begin{tabular}{@{}lcccccc@{}}
White Dwarf & Source & \teff (K) & log g & \masa_msun(1)
& \masa_msun(2) & \masa_msun(3) \\
0316 + 345   & BSL & 14880  & 7.61 & 0.40 & 0.414   & 0.452 \\
0339 + 523   & BSL & 13350  & 7.47 & 0.34 & 0.354   & 0.395 \\
0710 + 741   & BSL & 18930  & 7.45 & 0.35 & 0.377   & 0.418 \\
0957 - 666   & BRB & 27047  & 7.285 & 0.335 & 0.371 & 0.417 \\
1022 + 050   & BRB & 14481  & 7.483 & 0.351 & 0.364 & 0.405 \\
1101 + 364   & BSL & 13610  & 7.38 & 0.31 & 0.325   & 0.369 \\
1241 + 010   & BSL & 24010  & 7.22 & 0.31 & 0.342   & 0.390 \\
1317 + 453   & BSL & 14000  & 7.43 & 0.33 & 0.343   & 0.385 \\
1353 + 409   & BSL & 23580  & 7.54 & 0.40 & 0.427   & 0.470 \\
1614 + 136   & BSL & 22430  & 7.34 & 0.33 & 0.361   & 0.409 \\
1713 + 332   & BSL & 22030  & 7.40 & 0.35 & 0.376   & 0.420 \\
1824 + 040   & BRB & 14795  & 7.608 & 0.394 & 0.411 & 0.451 \\
2032 + 188   & BSL & 18540  & 7.48 & 0.36 & 0.385   & 0.425 \\
2331 + 290   & BSL & 27830  & 7.50 & 0.39 & 0.431   & 0.479 \\
2337 - 760   & BRB & 14295  & 7.507 & 0.354 & 0.372 & 0.410
\end{tabular}
\medskip

\teff, the
surface gravity (g) and the stellar mass (1) of the objects  were
taken  from  Bergeron,  Saffer   \&  Liebert  (BSL)  (1992)   and
Bragaglia,  Renzini  \&  Bergeron  (BRB)  (1995). The next column
(2) gives the stellar mass according to our helium - core models
without a hydrogen envelope and the last column lists the
stellar mass according to our helium - core models with a
hydrogen envelope of $M_{\rm H}/M=  10^{-3}$.
\end{minipage}
\end{table*}

Our main motivation for the publication of a detailed set of  low
- mass white dwarf models  like the one presented here  is, apart
from the fact that little attention has been paid in the past  to
the study of this kind of objects, that both finite - temperature
effects and hydrogen  envelopes substantially modify  the surface
gravity values of  zero - temperature,  helium - core -
degenerate
configurations. These features may turn out to be very  important
for the  interpretation of  the recent  and future  observational
data about low  - mass white  dwarfs. We think  that the detailed
low - mass models we  computed here by employing the  full scheme
of stellar evolution theory may help such an endeavour.

\begin{figure}
\epsfxsize=240pt
\begin{displaymath}
\epsfbox{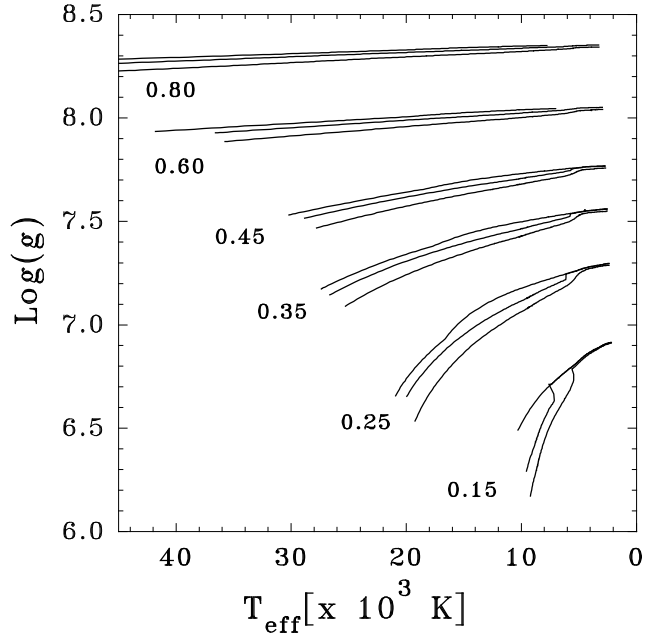}
\end{displaymath}
\caption{Surface gravities versus \teff
for selected white dwarf  models with \masa_msun=  0.15, 0.25,
0.35, 0.45, 0.60 and  0.80 and different hydrogen  envelopes. For
each stellar mass and from top to bottom the curves correspond to
sequences  having  hydrogen  envelopes  with  fractional  mass of
$M_{\rm H}/M$= 0 (no hydrogen envelope), $10^{-6}$ and $10^{-4}$,
respectively.  Note  that  in  the  case  of less massive models,
hydrogen envelopes appreciably reduce the surface gravity  values
of pure helium models.}
\label{fig_5}
\end{figure}

\begin{figure}
\epsfxsize=240pt
\begin{displaymath}
\epsfbox{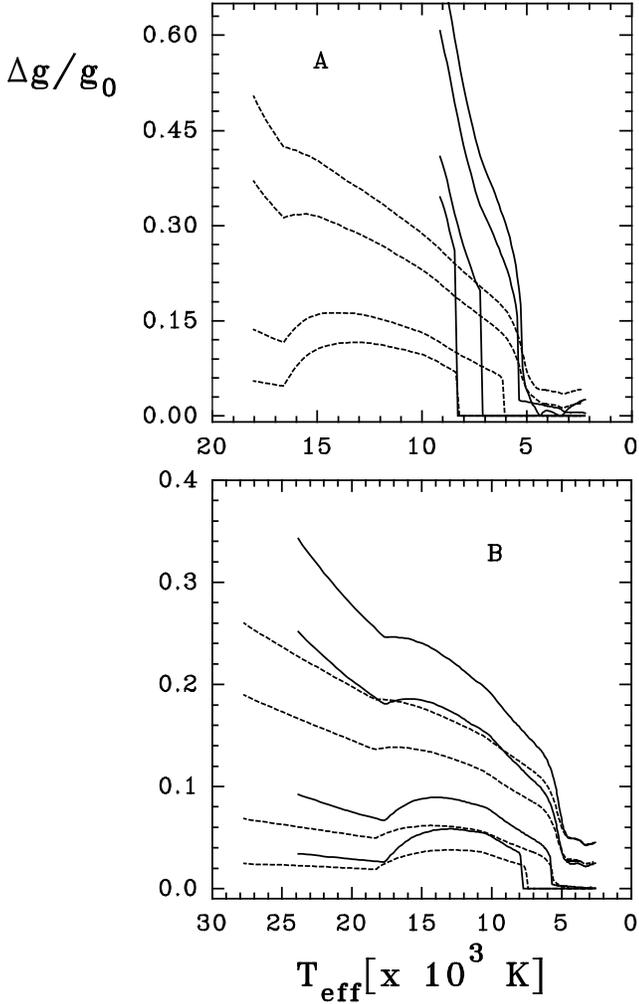}
\end{displaymath}
\caption{  (a):   Ratio   of  the
difference in $g$  values between low  - mass white  dwarf models
without  ($g_o$)  and  with  hydrogen  envelopes  to $g_o$ versus
\teff for models  with \masa_msun = 0.15  (solid lines)
and 0.25 (dashed  lines). For each  stellar mass and  from top to
bottom the curves correspond to sequences with hydrogen envelopes
of  fractional  mass  of  $M_{\rm  H}/M$=  $4  \times   10^{-4}$,
$10^{-4}$,  $10^{-6}$  and  $10^{-8}$,  respectively.
(b): As (a) but for models with with \masa_msun = 0.35
(solid lines) and 0.45  (dashed lines). Note that  thick hydrogen
envelopes  appreciably  reduce  the  $g$  values  of  pure helium
models. Note also the effect of convective mixing at low
\teff  on  models  with  thin  hydrogen envelopes.} \label{fig_6}
\end{figure}

In the  context of  the foregoing  paragraph we  show in  Figure~
\ref{fig_5} the surface  gravity $g$ (in  cgs units) in  terms of
\teff for some selected models. The effects of finite temperature
are clearly noticeable, particularly for less massive models.  As
is  well  known,  at  a  given  \teff  more  massive  models  are
characterized  by  smaller  radii.  As  \teff decreases the model
radius (gravity) gradually  becomes smaller (larger),  ultimately
reaching  an  almost  constant  value  as expected for a strongly
degenerate configuration,  in which  the mechanical  structure is
determined mainly by degenerate  electron pressure. As a  result,
stellar   parameters   asymptotically   reach   constant   values
corresponding to zero  temperature configurations. Note  also the
changes in the $g$ values brought about by rather thick  hydrogen
envelopes. Another observation  we can make  from this figure  is
that, at low \teff, convective mixing between hydrogen and helium
layers  increases  the  $g$  values  of models with thin hydrogen
envelopes. In  fact, convective  mixing changes  the outer  layer
composition from a hydrogen -  dominated to a helium -  dominated
one, thus giving rise to denser outer layers. From then on, their
subsequent  evolution  resembles  that  of  a  white  dwarf model
without   a   hydrogen   envelope,   as   can   be   noted   from
Figure~\ref{fig_5}  (see  also  Figure~\ref{fig_6}), particularly
for less massive models.

To clarify better  the role played  by hydrogen envelopes  in the
$g$ values of  low - mass  models, we show  in Figure~\ref{fig_6}
the reduction  in the  $g$ values  of pure  helium configurations
resulting from adding hydrogen envelopes of different  thickness.
More precisely, we plot in terms of \teff the quantity $\Delta  g
/  g_o  \equiv  (g_o-g_H)/g_o$  for  various  stellar  masses and
hydrogen envelopes ($g_o$ and $g_H$ stand for the surface gravity
of a helium - core configuration of a given stellar mass  without
and with  a hydrogen  envelope, respectively).  It is  clear that
thick hydrogen  envelopes appreciably  reduce the  $g$ values  of
pure helium models. At the low \teff of 15,000K for instance, the
$g$ values  of the  0.35 and  0.25 \msun  pure helium  models are
reduced, respectively,  by  20 and  30 per  cent if  a hydrogen
envelope of $M_{\rm H}/M= 10^{-4}$ is added. Such values increase
considerably at higher \teff. At \teff $\approx$ 17,000 K,  there
is a change in the slope of the curves stemming from the decrease
in the radiative opacity values after helium recombination.  This
causes  pure  helium  models  to  become  denser and hence to
have larger $g$ values.

\begin{figure}
\epsfxsize=240pt
\begin{displaymath}
\epsfbox{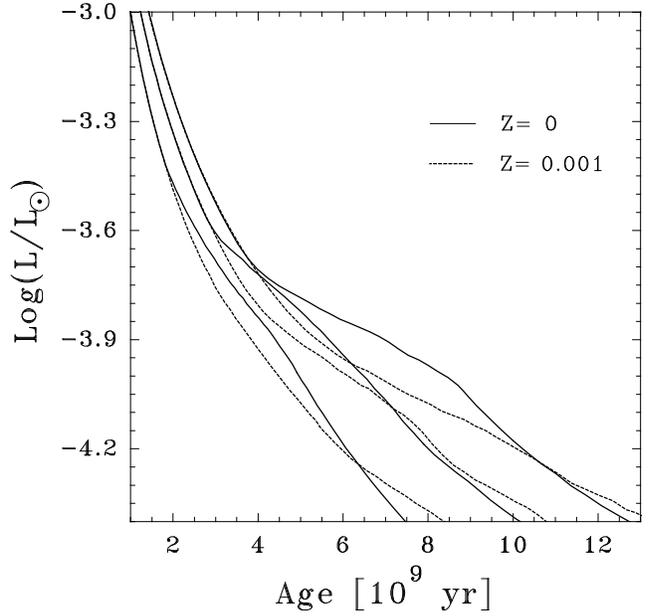}
\end{displaymath}
\caption{Surface luminosity versus age relation
for (from  top to  bottom) 0.40,  0.30 and  0.2 \msun helium
white  dwarf  models  with  $M_{\rm  H}/M$=  $10^{-4}$  and   for
metallicities $Z=$ 0 (dashed lines) and $Z=$ 0.001 (dotted
lines).
Note that at  high luminosities, cooling  is not affected  by the
assumed metallicity in the envelope.}
\label{fig_7}
\end{figure}

\begin{figure}
\epsfxsize=240pt
\begin{displaymath}
\epsfbox{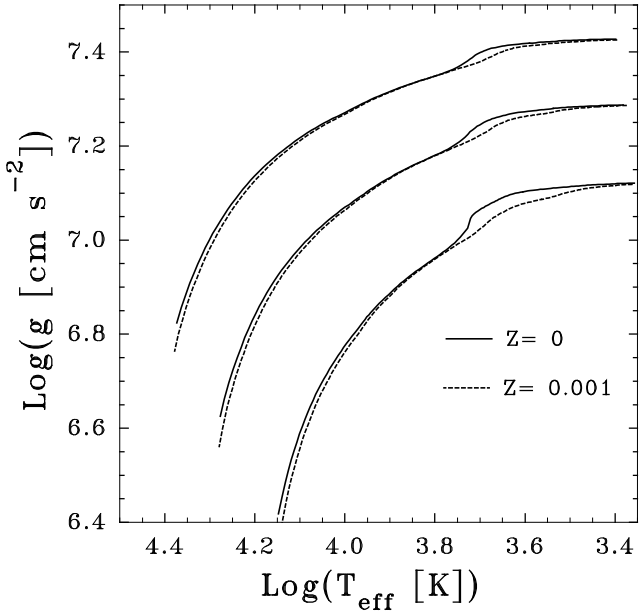}
\end{displaymath}
\caption{Surface gravities versus \teff
for (from top  to bottom) 0.30,  0.25 and 0.20  \msun helium
white  dwarf  models  with  $M_{\rm  H}/M$=  $10^{-4}$  and   for
metallicities  $Z=$  0  (solid  lines)  and  $Z=$  0.001  (dashed
lines).}
\label{fig_8}
\end{figure}

Non  -  negligible  differences  in  the structure and cooling of
white  dwarfs  may  also  arise  from the employment of different
metallicities in the  envelope, particularly at  low luminosities
where central temperature of models becomes strongly tied to the
details of the outer  layer chemical stratification (see  Tassoul
et al. 1990). This expectation is borne out by
Figure~\ref{fig_7},
in  which  we  compare  the  cooling  times of helium white dwarf
models for  two extreme  metallicities assumed  in the  envelope.
When convection  reaches the  domain of  degeneracy, the  central
temperature drops  substantially and  the star  has initially  an
excess of internal energy to  be radiated, thus giving rise  to a
lengthening of evolutionary times during that stage of evolution.
Because  models  with  lower  metallicities  are characterized by
deeper  convection  zones  (see  Figure~\ref{fig_1}),  this
effect
occurs at higher  luminosities in such  models and this  explains
their greater ages  as compared with  high - metallicity  models.
Eventually,  at  very  low  luminosities, more transparent models
evolve  more  rapidly,  as  expected.  We  have also analysed the
effect of metallicity on  surface gravity for helium  models (see
Figure~\ref{fig_8})  and we  found that  surface gravity  is
almost
insensitive to a specific choice of metallicity in the envelope.

Lastly, we have applied  our evolutionary models with  helium
cores
to  derive  stellar  masses  of  some  selected  low - mass white
dwarfs. To  this end,  we have picked  out low  surface gravity
white
dwarfs from the sample of  white dwarfs analysed by  Bergeron,
Saffer  \&  Liebert  (1992)  and  Bragaglia,  Renzini \& Bergeron
(1995), and  we  list  the  results  in  Table 3. The above cited
authors estimated the white dwarf masses from evolutionary models
with  pure  carbon  -   core  composition.  However,  using   our
evolutionary models with helium cores and no hydrogen envelope we
find   the   mass   values   to   be  appreciably underestimated,
particularly at  high temperatures.  It is  worth mentioning that
the objects listed  in Table 3  are white dwarfs  that most
likely have a hydrogen envelope. This being the case, the stellar
mass
should  be  estimated  from  evolutionary  models  with  hydrogen
envelopes.  As  shown  in  Table  3,  there  is  an   appreciable
difference  in  the  white  dwarf  mass when stellar masses are
derived from models with thick hydrogen envelopes ($M_{\rm  H}/M
\approx 10^{3}$).

In view of the preceding considerations, we conclude that
detailed
models of low - mass white  dwarfs such as presented in  this
study should be carefully taken into account, should the mass  of
a  white  dwarf  be  measured  by  applying the surface gravity -
\teff relation. This  is particularly true regarding  the
possibility of constraining the equation of state at neutron star
densities as inferred from observations of low - mass white dwarf
companions to millisecond  pulsars, such as  those studied by
van Kerkwijk et al. (1996).

Complete tables  containing the  results of  our calculations are
available           at           the   World Wide Web       site
http://www.fcaglp.unlp.edu.ar/$\sim$althaus/.          Additional
evolutionary sequences are obtained  upon request to the  authors
at their  e-mail address.  Features such  as surface  luminosity,
\teff, central density and temperature, surface  gravity,
stellar radius, age and hydrogen surface abundance are listed  in
the tables.

\section*{acknowledgments}

We are indebted to F. D'Antona for sending us the starting
model,
and to F. Rogers for providing us with his radiative opacity data
tables. We also thank T. Guillot and D. Saumon for their help  in
making available  to us  the low  - density  equation of state
that we have
employed here. We are also grateful to V. M. Canuto for providing
us
with the Canuto Goldman \& Mazzittelli model before publication.
We  also  thank  our  anonymous  referee,  whose comments greatly
improved the original version of this work.

{}

\end{document}